
\magnification=1200
\parindent 1.0truecm
\baselineskip=16pt
\rm
\null

\footline={\hfil}
\vglue 0.8truecm
\rightline{\bf DFUPG--51--92}
\rightline{\sl March 1992}
\vglue 2.0truecm
\centerline{\bf Low--Energy Kaon Interactions and Scattering on Nucleons }
\centerline{\bf and Light Nuclei}
\vglue 1.0truecm
\centerline{\sl Paolo M. Gensini }
\centerline{\sl Dip. di Fisica dell\'\ Universit\`a di Perugia, Perugia,
Italy, and }
\centerline{\sl I.N.F.N., Sezione di Perugia, Italy }
\vglue 6.0truecm
\centerline{ A chapter to be included in }
\centerline{\sl The DA$\mit\Phi$NE Physics Handbook }
\centerline{ (by the DA$\Phi$NE Theory Working Group) }
\centerline{ ed. by L. Maiani, G. Pancheri and N. Paver (INFN, Frascati 1992).}
\pageno=0
\vfill
\eject

\footline={\hss\tenrm\folio\hss}
\vglue 0.5truecm
\centerline{\bf Low--Energy Kaon Interactions and Scattering on Nucleons }
\centerline{\bf and Light Nuclei }
\vglue 1.0truecm
\centerline{\sl Paolo M. Gensini }
\centerline{\sl Dip. di Fisica dell\'\ Universit\`a di Perugia, Perugia,
Italy, and }
\centerline{\sl I.N.F.N., Sezione di Perugia, Italy }
\vglue 2.0truecm

We present in this contribution the basic formul\ae\ for the analysis of
low--momentum charged-- and neutral--kaon interactions in hydrogen, including
as well a (brief) description of the problems left open by past experiments
and of the improvements DA$\Phi$NE can be expected to offer over them.
Interactions in deuterium and other light nuclei will be only briefly
mentioned, and only in those respects touching directly upon the more
``elementary'' aspects of kaon--nucleon interactions.

\vglue 0.6truecm
\leftline{\bf 1. Introduction. }
\vglue 0.3truecm

DA$\Phi$NE is expected to produce, with ``standard'' assumptions about
luminosity and cross sections, about $1.25 \times 10^{10}\ K^\pm$'s and
$8.5 \times 10^9\ K_L$'s per year of operation, considering a conventional
``Snowmass year'' of $10^7\ s$. With a detector of KLOE's size one can thus
expect to observe up to millions of interactions per year,
even in the light element (quite
probably gaseous helium at atmospheric pressure) filling its fiducial volume.

A first question has thus to be answered: do these events contain useful
physics to be worth recording and interpreting? One could even go further
and envisage using DA$\Phi$NE as a source of high--resolution ($\Delta p/p$
of respectively $1.1 \times 10^{-2}$ for $K^\pm$'s and $1.5 \times 10^{-2}$
for $K_L$'s), low--momentum kaons ($127\ MeV/c$ for the former,
$110\ MeV/c$ for the
latter), to measure with a {\sl dedicated} detector
$KN$ and $\bar KN$ interactions in a toroidal volume
filled with gaseous $H_2$ or $D_2$, and possibly also interactions in heavier,
gaseous elements.

The machine has thus the ability to explore a kinematical
region very little investigated in the past:
only few bubble--chamber $K^-$ experiments in hydrogen (and deuterium)$^{1,2}$
plus very few data points on $K_S$ regeneration$^3$ exist, all with
extremely low statistics and more than a decade old (the last experiment$^2$
to cover this region was carried out by the TST Collaboration at the hydrogen
bubble chamber in NIMROD's low--energy kaon beam in the second half of the
seventies). Since in dispersive calculations
of low--energy parameters for $KN$ interactions (couplings, scattering
lengths, sigma terms) the bulk of the uncertainties
comes from the integrals over the unphysical region, to
describe which one must extrapolate down in energy from that just above
threshold, DA$\Phi$NE can be expected to make substantial improvements
over our present knowledge of those parameters.

The following sections are therefore dedicated to
illustrating both the level and the limitations of present--day
information on low--energy kaon--nucleon
physics, spotlighting those points which still await being clarified,
and where DA$\Phi$NE can be expected to improve. Being the
phenomenology in this case more complex than in the (strictly related)
pion--nucleon one, we shall start almost from scratch.

We shall also take the liberty of not going into the details of models,
in particular for the spectroscopic classification of the
$J^P = {1\over2}^-$, $S$--wave resonance $\Lambda(1405)$: data are so scarce,
for the moment being, that any interpretation of such a state is to be
regarded as purely conjectural$^4$.

\vglue 0.6truecm
\leftline{\bf 2. Amplitude formalism for two--body $KN$ and $\bar KN$
interactions. }
\vglue 0.3truecm

Any $a_1(0^-,q) + B_1({1\over2}^+,p) \to a_2(0^-,q') + B_2({1\over2}^+,p')$
process is most economically described in the centre--of--mass (c.m.) frame
by two amplitudes, $G(w,\theta)$ and $H(w,\theta)$, when the T--matrix element
$T_{\alpha\beta}$ is expressed in terms of the two--component Pauli spinors
$\chi_\alpha$ and $\chi_\beta$ (respectively for the final and initial
${1\over2}^+$ baryons) as $T_{\alpha\beta} = \chi_\alpha^{\dag} \hbox{\bf T}
\chi_\beta$, where
$$ \hbox{\bf T} = G(w,\theta) \times \hbox{\bf I} + i H(w,\theta) \times
(\vec\sigma \cdot \hat n) \eqno(1) $$
and $\hat n$ defines the normal to the scattering plane$^5$.

These c.m. amplitudes have a simple expansion in the partial waves
$T_{\ell\pm}(w) = (\eta_{\ell\pm} e^{2 i \delta_{\ell\pm}} - 1) / 2 i q$,
given by
$$ G_N(w,\theta) = \sum_{\ell=0}^\infty [ (\ell + 1) T_{\ell+}(w) + \ell \
T_{\ell-}(w) ] P_\ell(\cos\theta) \eqno(2) $$
$$ H_N(w,\theta) = \sum_{\ell=1}^\infty [ T_{\ell+}(w) - T_{\ell-}(w) ]
P'_\ell(\cos\theta)\ , \eqno(3) $$
where the subscript $N$ indicates that only the {\sl purely nuclear} part of
the interaction has been considered.

To describe adequately the data, the amplitudes must also include
electromagnetism and can be rewritten as
$$ G(w,\theta) = \tilde G_N(w,\theta) + G_C(w,\theta) \eqno(4)$$
$$ H(w,\theta) = \tilde H_N(w,\theta) + H_C(w,\theta) \ ,\eqno(5)$$
where the tilded nuclear amplitudes differ from the untilded ones only in the
Coulomb shifts $\sigma_{\ell\pm}^{\rm in(fin)}$ (differing from zero only when
{\sl both} particles in the initial (final) state are charged) having been
applied to each partial wave $T_{\ell\pm}$, namely when
$$ T_{\ell\pm}\ \to\ \tilde T_{\ell\pm} = e^{i \sigma_{\ell\pm}^{\rm
in}} T_{\ell\pm}(w) e^{i \sigma_{\ell\pm}^{\rm fin}} \ . \eqno(6)$$

The one--photon--exchange amplitudes $G_C$ and $H_C$ (of course
absent for charge-- and/or
strangeness--exchange processes, but not for $K_S$ regeneration, which at $t
\neq 0$ goes also via one--photon exchange) can be
expressed in terms of the Dirac nucleon form factors as$^{5,6}$
$$ G_C(w,\theta) = \pm e^{\pm i \phi_C} \cdot \{ ( {{2 q \gamma} \over t}
+ {\alpha
\over {2 w}} {{w + m} \over {E + m}} ) \cdot F_1(t) + [ w - m + {t \over {4 (E
+ m)}} ] \cdot {{\alpha F_2(t)} \over {2 w m}} \} \cdot F_K(t) \eqno(7) $$
and
$$ H_C(w,\theta) = \pm {{\alpha F_K(t)} \over {2 w \tan{1\over2}\theta}} \cdot
\{ {{w + m} \over {E + m}} \cdot F_1(t) + [w + {t \over {4 (E + m)}}] \cdot
{{F_2(t)} \over m} \} \eqno(8) $$
for the interactions of (respectively) $K^\pm$ with nucleons, while for $K_S$
regeneration one has to change the sign of the isovector part of the kaon form
factor $F_K(t)$. Here
$\gamma = \alpha \cdot (w^2 - m^2 - \mu^2) / 2 q w$ and the Coulomb
phase $\phi_C$ is defined as
$$ \phi_C = - \gamma \log(\sin^2{1\over2}\theta) + \gamma \cdot
\int_{-4q^2}^0 {{dt} \over t} \cdot [1 - F_K(t) F_1(t)] \eqno(9) $$
for charged kaons scattering on protons, while it reduces to
$$ \phi_C = -\gamma \int_{-4q^2}^0 {{dt}\over t} F_K(t) F_1(t) \eqno(9') $$
for processes involving $K^0$'s and/or neutrons.

We have denoted with $w$ and $\theta$ respectively the total energy
and the scattering angle in the c.m. frame, $q = [{1\over4} w^2 - {1\over2}
(m^2 + \mu^2) + (m^2 - \mu^2)^2 / 4 w^2]^{1/2}$ is the c.m. momentum (in the
initial state: for inelastic processes, including charge exchange, we shall
indicate final--state kinematical quantities with primes), $E$ the total
energy of the baryon in the c.m. frame, $E = (w^2 + m^2 - \mu^2)/2w$, and $t$
the squared momentum--transfer, $t = m^2 + m'^2 - 2 E E' + 2 q q' \cos\theta$.
We shall also use the laboratory--frame, initial--meson momentum $k =
{1\over2} (\omega^2 - \mu^2)^{1/2}$ and energy $\omega$, related to the
c.m. total energy via $\omega = (w^2 - m^2 - \mu^2)/2m$, and, besides $t$,
the two other
Mandelstam variables $s = w^2$ and $u$, the square of the c.m. total energy
for the crossed channel $\bar a_2(0^-) + B_1({1\over2}^+) \to \bar a_1(0^-)
+ B_2({1\over2}^+)$, obeying {\sl on the mass shell} the indentity $s
+ t + u = m^2 + m'^2 + \mu^2 + \mu'^2$.

The shifts $\sigma_{\ell\pm}$ have been computed accurately by Tromborg,
Waldenstr\"om and \O verb\o$^6$ in a dispersive formalism
for the $\pi N$ case:
the same formalism$^7$
could be extended (but has not been up to now) also to the
$K N$ and $\bar K N$ ones. Minor corrections remain to be applied
to the phases $\delta_{\ell\pm}$ and to the elasticities $\eta_{\ell\pm}$, to
extract their purely nuclear parts$^5$; a way of removing them efficiently
has been devised by the Karlsruhe--Helsinki group for $\pi N$ amplitudes:
it consists in starting from a (preliminary) set of phase shifts, calculating
 from them the corrections in the above--mentioned dispersive
formalism$^{6,7}$, then the changes in the observables brought about by these
latter, and finally correcting the data for these effects and starting
the phase--shift analysis all over again, this time from the ``corrected
data''. This procedure has turned out to be both self--consistent and
fast$^5$.

In terms of the amplitudes $G$ and $H$ the c.m. differential cross
sections for an unpolarized target (which will most surely be the case in
DA$\Phi$NE's almost $4\pi$ geometry) take the simple form
$$ {{d\sigma} \over {d\Omega}} = {1\over2} \sum_{\alpha,\beta} \vert
T_{\alpha\beta} \vert^2 = \vert G \vert^2 + \vert H \vert^2 \ . \eqno(10) $$

The other observables possibly accessible at DA$\Phi$NE, in the
strangeness--exchange processes $\bar K N \rightarrow \pi \Lambda$ and
$\bar K N \rightarrow \pi \Sigma$, are the polarizations $P_Y$ ($Y = \Lambda$
or $\Sigma$) of the final hyperons,
measurable through the asymmetries $\alpha$ of their weak nonleptonic decays
$\Lambda \rightarrow \pi^- p$ or $\pi^0 n$, for both of
which we have an asymmetry $\alpha \simeq 0.64$, and $\Sigma^+ \rightarrow
\pi^0 p$ for which the asymmetry is $\alpha \simeq - 0.98$, while there is
very little chance to be able to use the neutron decay modes $\Sigma^\pm
\rightarrow \pi^\pm n$, which have the very small asymmetries $\alpha \simeq
\pm 0.068$; we have for these quantities
$$ P_Y \cdot ({{d\sigma} \over {d\Omega}}) = 2\ \hbox{\rm Im} \ (GH^*) \ .
\eqno(11) $$

Note that, for an $(S+P)$--wave parametrization (fully adequate at such low
momenta), while the {\sl integrated} cross sections depend only {\sl
quadratically} on the $P$--waves, both the first Legendre coefficients of the
differential cross sections
$$ L_1 = {1\over2} \int_{-1}^{+1} \cos\theta\ ({{d\sigma} \over {d\Omega}})\
d\cos\theta = {2\over3}\ \hbox{\rm Re}\ [T_{0+} (2 T_{1+} + T_{1-})^* + \dots
] \eqno(12)$$
and the polarizations
$$ P_Y \cdot ({{d\sigma} \over {d\Omega}}) = 2\ \hbox{\rm Im}\ [T_{0+} (T_{1+}
- T_{1-})^* + 3 T_{1-} T_{1+}^* \cos\theta + \dots ] \sin\theta \eqno(13) $$
are essentially {\sl linear} in the small $P$--wave contributions, and give
complementary information on these latter. It is perhaps not useless to remind
the reader that the low statistics of the experiments, performed {\sl only} up
to the late seventies, have not been enough to determine any of these
parameters, putting only rather generous (and utterly useless) upper
bounds$^2$ on $\vert L_1 \vert$ for the hyperon production channels.

We shall now devote the last part of this section to show explicitly why
this absence of {\sl direct} information on the low--energy behaviour of the
$P$--waves has been a serious shortcoming for $\bar K N$ amplitude analyses.
Remember that we know, from {\sl production} experiments, that the $I = 1$, $S
= -1$ $T_{1+}$ partial wave resonates {\sl below} threshold at a c.m.
energy around $w = 1385\ MeV$, the mass of the isovector member of the $J^P =
{3\over2}^+$ decuplet.

One has to turn from the Pauli amplitudes $G$ and $H$ to the invariant
amplitudes $A(s,t)$ and $B(s,t)$, defined in term of four--component Dirac
spinors as
$$ 2\pi w \ T_{\alpha\beta} = \bar u_\alpha(p') [A(s,t) + B(s,t)\cdot
\gamma^\mu Q_\mu ] u_\beta(p) \ , \eqno(14) $$
where $Q = {1\over2} (q + q')$, the average between incoming-- and
outgoing--meson c.m. four--momenta: these amplitudes obey simple
crossing relations and are free of kinematical
singularities, so that they are the ones to be used, rather than $G$ and $H$,
for any analytical extrapolation purpose; it is also customary to use the
combination $D(\nu,t) = A(\nu,t) + \nu \cdot B(\nu,t)$, where $\nu = (s - u)
/ 2(mm')^{1/2}$, which has the same properties as $A(\nu,t)$ under crossing,
and furthermore, for elastic scattering, obeys the optical theorem
in the simple form
$$ \hbox{\rm Im} \ D(\nu,t=0) = k \cdot \sigma_{tot} \ , \eqno(15) $$
where of course all electromagnetic effects must be subtracted on {\sl both}
sides.

One can rewrite $A$ and $B$ in terms of $G$ and $H$, and thus reexpress them
through the partial waves $T_{\ell\pm}$, by projecting eq. (14)
on the different spin states (polarized {\sl perpendicularly} to the
scattering plane) and obtain in the most general kinematics

$$ A(\nu,t) = {{4 \pi} \over {(E + m)^{1/2} (E' + m')^{1/2}}} \{ [w +
{1\over2} (m + m')] G(w,\theta) + $$
$$+ [(E + m) (E' + m') \{ w - {1\over2} (m + m') \} + \{ {1\over2} t + E E' -
{1\over2} (m^2 + m'^2) \} \{ w + {1\over2} (m + m') \} ] \cdot$$
$$\cdot {{H(w,\theta)} \over {q q' \sin\theta}} \} \ , \eqno(16)$$
and
$$ B(\nu,t) = {{4 \pi} \over {(E + m)^{1/2} (E' + m')^{1/2}}} \{ G(w,\theta)
- $$
$$- [(E + m) (E' + m') - {1\over2} t - E E' + {1\over2} (m^2 + m'^2)]
{{H(w,\theta)} \over {q q' \sin\theta}} \} \ . \eqno(17)$$
\vglue 0.3truecm

Considering for sake of simplicity forward elastic scattering only, the
amplitudes become, leaving out $D$-- and higher waves,
$$ D(\nu,0) = {{4 \pi w} \over m} [T_{0+} + 2 T_{1+} + T_{1-} + \dots ]
\eqno(18) $$
and
$$ B(\nu,0) = {{4 \pi w} \over {m q^2}} [(E - m) T_{0+} - 2 (2m - E) T_{1+} +
(E + m) T_{1-} + \dots ] \ ; \eqno(19) $$
introducing the (complex) scattering lengths $a_{\ell\pm}$ and (complex)
effective ranges $r_{\ell\pm}$ one can expand up to $O(q^2)$ the partial
waves close to threshold, and obtain for the forward $D$ amplitudes
$$ D(q,0) = 4 \pi (1 + {\mu \over m}) \{ a_{0+} + i a_{0+}^2 q + [2 a_{1+} +
a_{1-} - (a_{0+} + {1\over2} r_{0+}) a_{0+}^2 - {a_{0+} \over {2 m \mu}}] q^2
+ \dots \} \ , \eqno(20) $$
dominated by the $S$--waves, while for the $B$ amplitudes the same
approximations give
$$ B(q,0) = {{2 \pi} \over m} (1 + {\mu \over m}) [a_{0+} - 4 m^2 (a_{1+} -
a_{1-}) + i a_{0+}^2 q + \dots] \ , \eqno(21) $$
where the factor $4 m^2 \simeq 90\ fm^{-2}$ enhances considerably the
contributions by the low--energy P--waves (virtually unkown), rendering
practically useless the unsubtracted dispersion relation for the
better converging $B$ amplitudes, so important for the $\pi N$ case
in fixing accurately the values
of the coupling constant $f^2$ and of the S--wave scattering lengths$^5$.

\vglue 0.6truecm
\leftline{\bf 3. Open channels and baryon spectroscopy at DA$\Phi$NE. }
\vglue 0.3truecm

As mentioned above, in the momentum region which could be explored by the
kaons coming from the decays of a $\phi$--resonance formed at rest in an $e^+
e^-$ collision, we have only data from low--statistics experiments, mostly
hydrogen bubble--chamber ones on $K^- p$ (and $K^-d$) interactions$^{1,2}$
(dating from the early sixties trough the late seventies),
plus scant data from $K_L$
interactions and $K_S$ regeneration, mostly on hydrogen$^3$.

The channels, open at a laboratory energy $\omega = {1\over2} M_\phi$ (for
$K^\pm$'s to obtain the exact value of $\omega$ one has to include their
energy losses through ionization as well), are tabulated below for
interactions with {\sl free} protons and neutrons, together with their
threshold energies $E_{thr}$ (in $MeV$), strangeness and isospin(s).
We do not list
$K^+$--initiated processes, which are (apart from charge exchange) purely
elastic in this energy region.

\vglue 1.0truecm
\centerline{\bf Table I}
\vglue 0.3truecm
\hrule
\vglue 0.3truecm
$$\vbox{\halign{#\hfil&\qquad\qquad\hfil#\hfil&\qquad\qquad\hfil#\hfil&\qquad
\qquad\hfil#\hfil\cr
\qquad \sl Channel&$E_{thr}/MeV$&$S$&$I$\cr
$K^-p, K^0_Ln\ \rightarrow\ \pi^0\Lambda$&$1250.6$&--1&1\cr
$K^-p, K^0_Ln\ \rightarrow\ \pi^0\Sigma^0$&$1327.5$&--1&0\cr
$K^-p, K^0_Ln\ \rightarrow\ \pi^-\Sigma^+$&$1328.9$&--1&0,1\cr
$K^-p, K^0_Ln\ \rightarrow\ \pi^+\Sigma^-$&$1337.0$&--1&0,1\cr
$K^-p, K^0_Ln\ \rightarrow\ \pi^0\pi^0\Lambda$&$1385.6$&--1&0\cr
$K^-p, K^0_Ln\ \rightarrow\ \pi^+\pi^-\Lambda$&$1394.8$&--1&0,1\cr
$K^-p, K^0_Ln\ \rightarrow\ K^-p$&$1431.9$&--1&0,1\cr
$K^-p, K^0_Ln\ \rightarrow\ K^0_Sn$&$1437.2$&--1&0,1\cr
\qquad''\qquad''&''&+1&1$^{\dag}$\cr}}$$
\hrule
\vglue 0.3truecm
\centerline{\dag) This amplitude only appears in the regeneration process
$K^0_L n
\rightarrow K^0_S n$.}

\vglue 1.5truecm
\centerline{\bf Table II}
\vglue 0.3truecm
\hrule
\vglue 0.3truecm
$$\vbox{\halign{#\hfil&\qquad\qquad\hfil#\hfil&\qquad\qquad\hfil#\hfil&\qquad
\qquad\hfil#\hfil\cr
\quad\sl Channel&$E_{thr}/MeV$&$S$&$I$\cr
$K^-n\ \rightarrow\ \pi^-\Lambda$&$1255.2$&--1&1\cr
$K^-n\ \rightarrow\ \pi^-\Sigma^0$&$1332.1$&--1&1\cr
$K^-n\ \rightarrow\ \pi^0\Sigma^-$&$1332.1$&--1&1\cr
$K^-n\ \rightarrow\ \pi^0\pi^-\Lambda$&$1388.2$&--1&1\cr
$K^-n\ \rightarrow\ K^-n$&$1433.2$&--1&1\cr}}$$
\hrule\vfill\eject

\centerline{\bf Table III}
\vglue 0.3truecm
\hrule
\vglue 0.3truecm
$$\vbox{\halign{#\hfil&\qquad\qquad\hfil#\hfil&\qquad\qquad\hfil#\hfil&\qquad
\qquad\hfil#\hfil\cr
\quad \sl Channel&$E_{thr}/MeV$&$S$&$I$\cr
$K^0_Lp\ \rightarrow\ \pi^+\Lambda$&$1255.2$&--1&1\cr
$K^0_Lp\ \rightarrow\ \pi^0\Sigma^+$&$1324.3$&--1&1\cr
$K^0_Lp\ \rightarrow\ \pi^+\Sigma^0$&$1332.1$&--1&1\cr
$K^0_Lp\ \rightarrow\ \pi^0\pi^+\Lambda$&$1388.2$&--1&1\cr
$K^0_Lp\ \rightarrow\ K^+n$&$1433.2$&+1&0,1\cr
$K^0_Lp\ \rightarrow\ K^0_Sp$&$1435.9$&+1&0,1\cr
\quad''\qquad ''&''&--1&1\cr}}$$
\hrule
\vglue 1.0truecm

For interactions in hydrogen, the c.m. energy available for each final
state is limited by momentum conservation to the initial total c.m. energy,
equal (neglecting energy losses) to $w = (m_p^2 + \mu_K^2 + m_p
M_\phi)^{1/2}$, or $1442.4\ MeV$ for incident $K^\pm$'s and $1443.8\ MeV$
for incident $K_L$'s. Energy losses for charged kaons can be exploited
(using the inner parts of the detector as a ``moderator'') to explore $K^- p$
interactions in a {\sl limited} momentum range, down to the charge--exchange
threshold at $w = 1437.2\ MeV$, corresponding to a $K^-$ laboratory momentum
of about $90\ MeV/c$.

For interactions in deuterium (or in heavier nuclei), momentum can be carried
away by ``spectator'' nucleons, and one can explore
each inelastic channel from the highest available energy down to threshold.
The possibility of reaching energies below the $\bar K N$ threshold is
particularly desirable, since the $\bar K N$ unphysical region contains two
resonances, the $I = 0$, $S$--wave $\Lambda(1405)$ and the $I = 1$, $J^P =
{3\over2}^+$ $P$--wave $\Sigma(1385)$, observed mostly in production
experiments (and, in the first case, in very limited statistics ones$^8$),
so that
the information on their couplings to the $\bar K N$ channel relies {\sl
entirely} on extrapolations below threshold of the analyses of the low--energy
data. The coupling of the $\Sigma(1385)$ to the $\bar K N$ channel, for
instance, can at present be determined only via forward dispersion relations
involving the {\sl total sum} of data collected at $t \simeq 0$, but with
uncertainties which are, {\sl at their best}, still of the order of 50\% of
the value expected from flavour--$SU(3)$ symmetry$^9$; as for the
$\Lambda(1405)$, even its spectroscopic classification is still an open
problem, {\sl vis--\`a--vis} the paucity and (lack of)
quality of the {\sl best available} data$^{4,10}$.

A {\sl formation} experiment on {\sl bound} nucleons in an (almost) $4\pi$
apparatus with good efficiency and resolution for low--momentum $\gamma$'s
(such as KLOE) can reconstruct and measure a channel such as $K^-p \rightarrow
\pi^0\Sigma^0$ (only {\sl above} the $\bar KN$ threshold) or $K^-d\rightarrow
\pi^0\Sigma^0 n_s$ (both {\sl above} and {\sl below} threshold),
which is pure $I = 0$: up to now all analyses on the $\Lambda(1405)$ have been
limited to charged channels$^8$, being forced to assume the $I = 1$
contamination in their samples to be either negligible or smooth and not
interfering with the resonance signal. This situation is particularly
unsatisfactory, in view of the fact that the various spectroscopic models
proposed for the classification of the $\Lambda(1405)$ differ mostly in the
detailed resonance {\sl shape}, rather than in its couplings: now, it is
precisely the shape which could be drastically changed even by a moderate
amount of interference with an $I = 1$ ``background''. Note also that, having
in the same apparatus and at almost the same energy {\sl tagged} $K^-$
and $K_L$ produced
at the same point, one can separate $I=0$ and $I=1$
channels with a minimum of systematic uncertainties, by
measuring all channels $K_L p \rightarrow \pi^0\Sigma^+$,
$\pi^+\Sigma^0$ and $K^- p \rightarrow \pi^-\Sigma^+$, $\pi^+\Sigma^-$,
besides, of course, the above--mentioned, pure $I=0$,
$K^- p \rightarrow \pi^0\Sigma^0$ one.

Another class of inelastic processes, which are expected to be produced
(even if at a much smaller rate) by DA$\Phi$NE's $\bar K$'s, is
radiative capture, leading in hydrogen to the final states $\gamma\Lambda$
and $\gamma\Sigma^0$ for incident $K^-$'s, and, for incident $K_L$'s, to the
final state $\gamma\Sigma^+$: in deuterium, one expects to observe the capture
processes by neutrons,
$K^-d\rightarrow\gamma\Sigma^-p_s$ and $K_Ld\rightarrow\gamma
\Sigma^0p_s$, $\gamma\Lambda p_s$, as well.
Observation of these processes has been
limited up to now to searches for photons emitted after capture
of $K^-$'s stopped in liquid hydrogen (and deuterium): alas,
the spectra in these experiments are
dominated by photons from unreconstructed $\pi^0$ and $\Sigma^0$
decays$^{11}$. This poses serious difficulties already at the
level of separation of signals from backgrounds, since
(in $K^-p$ capture at rest) only the photon line from the first
final state falls
{\sl just above} the endpoint of the photons from
decays of the $\pi^0$'s in the
$\pi^0\Lambda$ final state, while that from the second falls right on top
of this latter: indeed, these experiments were able to produce,
within quite large errors, only an estimate of
the respective branching ratios.

The $4\pi$ geometry possible at DA$\Phi$NE, combined with the ``transparency''
of a KLOE--like apparatus, its high efficiency for photon detection and its
good resolution for spatial reconstruction of the events, should make possible
the full identification of the final states and therefore the measurement of
the absolute cross sections for these processes, although in flight and not
at rest.

This difference can be appreciated when comparing with
theoretical predictions:
the main contributions to radiative captures are commonly thought to
come from radiative decays of resonant levels in the $\bar KN$ system$^{12}$,
while the total hyperon production cross section is expected to come from
both resonant and non--resonant intermediate states. An estimate of the
branching ratios would therefore be quite sensitive to the latter, while a
prediction of the absolute cross sections should not.

Data$^{11}$ are presently indicating branching ratios around
$0.9 \times 10^{-3}$ for $K^-p\rightarrow\gamma\Lambda$
and $1.4\times 10^{-3}$ for $K^-p\rightarrow\gamma\Sigma^0$,
with errors of the order of 15\% on both rates: most
theoretical models$^{13}$ tend to give the first rate larger than the
second, with both values consistently higher than the observed ones. Only a
cloudy--bag--model estimate$^{14}$ exhibits the trend appearing
(although only at a $2\sigma$ level, and therefore waiting for confirmation
by better data) from the first experimental determinations, but this is
the only respect in which this model agrees with the data, still giving
branching ratios larger than observations by a factor two.

Data are also interpretable in terms of $\Lambda(1405)$ electromagnetic
transition moments$^{12}$: this interpretation of measurements taken at a
single energy, or over a limited interval, is clearly subject to the effect
of the interference between this state and all other contributions, such as
the $\Lambda$-- and $\Sigma$--hyperon poles and other resonant states such
as the $\Sigma(1385)$ and the $\Lambda(1520)$, not to mention $t$--channel
exchanges (since at least $K$--exchange has to be included, to ensure gauge
invariance of the Born approximation). An extraction of the $\Lambda(1405)$
moments, relatively freer of these uncertainties, requires measurements of
the final states $\gamma\Lambda$ and $\gamma\Sigma$ (if possible, in
different charge states) over the {\sl unphysical region},
using (gaseous) deuterium or helium as a ``target''.
Rates are expected to be only of the order of $10^4\ events/y$,
but it must be kept in mind that such a low rate corresponds already to
statistics even better than those of the {\sl best} experiment performed
till now on the
shape of the $\Lambda(1405)\rightarrow\pi\Sigma$ decay spectrum$^8$.

\vglue 0.6truecm
\leftline{\bf 4. The K-matrix (or M-matrix) formalism. }
\vglue 0.3truecm

An adequate description of the low--energy $\bar K N$ partial waves must
couple at least the dominant, two--body inelastic channels to each other and
to the elastic one; the three--body channel $\pi\pi\Lambda$ is expected to be
suppressed, for $J^P = {1\over2}^-$, by the angular momentum barrier, but it
could contribute appreciably to the $I=0$, $J^P={1\over2}^+$ $P$--wave, due
to the strong final--state interaction of two pions in an $I=0$ $S$--wave.
Note that {\sl most} bubble chamber experiments were unable to fully
reconstruct the events at the lowest momenta,
and therefore often assumed all {\sl
directly} produced $\Lambda$'s to come from the $\pi\Lambda$ channel {\sl
alone}, neglecting altogether the small $\pi\pi\Lambda$ contribution.

The appropriate formalism is to introduce a K--matrix description (sometimes
it is convenient to use, instead of the K--matrix, its inverse, also known as
the M--matrix), defined in the isospin eigenchannel notation as
$$ \hbox{\bf K}_{\ell\pm}^{-1} = \hbox{\bf M}_{\ell\pm} =
\hbox{\bf T}_{\ell\pm}^{-1} + i\ \hbox{\bf Q}^{2\ell +1}\ , \eqno(22)$$
for both $I=0, 1$ $S$--waves (and perhaps also for the four $P$--waves
as well). The K--matrices, assuming $SU(2)$
symmetry, describe the $S$--wave data at a given energy in terms of {\sl nine}
real parameters (six for $I=1$ and three for $I=0$), while the experimentally
accessible processes can be described, assuming pure $S$--waves in the
same symmetry limit, by only {\sl
six} independent parameters, which can be chosen to be the two (complex)
amplitudes $A_0,\ A_1$ for the $\bar K N \rightarrow \bar K N$ channel, the
phase difference $\phi$ between the $I=0$ and $I=1$ $\pi\Sigma$ production
amplitudes, and the ratio $\epsilon$ between the $\pi\Lambda$ production cross
section and that for total hyperon production in an $I=1$ state$^{15}$.

Thus a single--energy measurement does not allow a complete determination of
the K--matrix elements at that energy. Using high--statistics measurements
at different momenta, and assuming either {\sl constant} K--matrices or (if
more complexity were needed) {\sl effective--range} M--matrices could
{\sl in principle} fully determine the matrix elements:
but for this to be possible one has to be able to subtract out
the (small) $P$--wave contributions to the integrated cross sections
$$ \sigma = 4\pi L_0 = 2\pi \int_{-1}^{+1} ({{d\sigma}\over{d\Omega}})\
d\cos\theta = 4\pi [ \vert T_{0+} \vert^2 + 2 \vert T_{1+}\vert^2 + \vert
T_{1-}\vert^2 + \dots ] \ , \eqno(23)$$
which could be obtained either from $L_1$ alone, for
the elastic and charge--exchange channels, or from both $L_1$ and $P_Y$,
which give {\sl complementary} information, for the hyperon production
channels. None of these quantities has been
measured up to now: the TST
Collaboration tried to extract $L_1$ from some of their low--statistics data,
but found results consistent with zero within their obviously very large
errors$^2$. At the same level of accuracy, one should also be able to isolate
and separate out the $\pi\pi\Lambda$ channel contribution as well.

Remember that an accurate analysis has also to
include the {\sl complete} e.m.
correct-ions$^{6,7}$: up to now all $\bar KN$ analyses have relied on the
old, approximate formul\ae\ derived by Dalitz and Tuan for a pure $S$--wave
scattering$^{16}$.

To fix the {\sl redundant} K--matrix parameters,
different authors have tried different methods:
some have used the data on the shape of the $\pi \Sigma$ spectrum from
production experiments$^{17}$,
others have constrained the amplitudes in the unphysical region by
imposing consistency with dispersion relations for the amplitudes $D$
for both $K^\pm p$ and $K^\pm n$ forward elastic scattering$^{18-20}$,
relying on the existence of
accurate data on the total cross sections at higher energies. More
recently, some attempts have been made to combine both constraints into a
``global'' analysis, but with no better results than each of them taken
separately$^{21}$.

Unfortunately, neither of these methods has been very powerful,
because of the low
statistics of the $\pi\Sigma$ production data on one side, and on the other
because of the need to use for the dispersion relations
the often not very accurate
information (and particularly so for the $K^\pm n$ amplitudes) on the
real--to--imaginary--part ratios.

We list below (without errors, often meaningless since
the parameters are strongly correlated, and therefore not even quoted
by some of the authors) the constant K--matrices
found by Chao {\sl et al.} using the first method$^{17}$
(which {\sl did not} include the TST Collaboration data),
and the more complex parametrization found by A.D. Martin using the
second$^{19,20}$ (and including the {\sl preliminary}
TST data). Note that to describe the data for $I=0$ both {\sl above} and {\sl
below} threshold A.D. Martin was forced to introduce a
``constant--effective--range''
M--matrix, where ${\bf M}^{(0)} = ({\bf K}^{(0)})^{-1} = {\bf A} + {\bf R}
k^2$, with three more ``effective range'' parameters, so that
to make the two analyses comparable we list separately his {\sl
threshold} K--matrix values.

The purpose of this table is to show that there is considerable uncertainty
even on the value of the $K^{(I)}_{NN}$ elements of the K--matrices
(the real parts of
the corresponding scattering lengths): the data have been
re--analyzed by Dalitz {\sl et al.}$^{21}$, using
{\sl both} sets of constraints with
different weigths and different parametrizations, and yielding
a variety of fits,
all of them of about the same overall quality and
none of them improving very
much over the above ones.

Just to highlight the difficulties met in describing the data (probably
plagued by inconsistencies between different experiments and BY
large systematic uncertainties), we point out
that A.D. Martin himself$^{19}$ found that including in his analysis a
$\Sigma(1385)$ resonance at the right position, with the width given by
the production experiments (and listed in the Particle Data Group tables)
and the coupling to the $\bar K N$ channel dictated by flavour--$SU(3)$
symmetry, was worsening rather than improving the fits obtained
{\sl neglecting} it altogether: his analysis therefore considers
the $\Sigma$ Born--term contribution a ``superposition'' of the former and
of that of the $P$--wave resonance, a rather unsavoury situation considering
the different $J^P$ quantum numbers of the two states, which may raise
questions about the applicability of his analysis away from $t \simeq 0$.
Note that a similar superposition
has to be considered in the $K^\pm p$ dispersion relations for the $\Sigma$--
and $\Lambda$--pole contributions, which can not be separated from each other,
but here the two states contribute to the same partial wave, and the
$\Sigma$--pole can be extracted independently from $K^\pm n$ scattering
(or $K_S$ regeration on protons) data$^{22}$.

\vglue 0.8truecm
\centerline{\bf Table IV}
\vglue 0.3truecm
\hrule
$$\vbox{\halign{\hfil # \hfil & \quad \hfil # \hfil & \quad \hfil # \hfil
& \quad \hfil # \hfil \cr
\sl Chao et al. & & \sl A. D. Martin & \cr
& & & \cr
$K_{NN}^{(0)} = -1.56 fm$ & $A_{NN} = -0.07 fm^{-1}$ & $R_{NN} = +0.18 fm$
& $K_{NN}^{(0)}(0) = -1.65 fm$ \cr
$K_{N\Sigma}^{(0)} = -0.92 fm$ & $A_{N\Sigma} = -1.02 fm^{-1}$ & $R_{N\Sigma}
= +0.19 fm$ & $K_{N\Sigma}^{(0)}(0) = +0.16 fm$ \cr
$K_{\Sigma\Sigma}^{(0)} = +0.07 fm$ & $A_{\Sigma\Sigma} = +1.94 fm^{-1}$ &
$R_{\Sigma\Sigma} = -1.09 fm$ & $K_{\Sigma\Sigma}^{(0)}(0) = -0.15 fm$ \cr
& & & \cr
$K_{NN}^{(1)} = +0.76 fm$ & & &$K_{NN}^{(1)} = +1.07 fm$ \cr
$K_{N\Sigma}^{(1)} = -0.97 fm$ & & & $K_{N\Sigma}^{(1)} = -1.32 fm$ \cr
$K_{N\Lambda}^{(1)} = -0.66 fm$ & & & $K_{N\Lambda}^{(1)} = -0.30 fm$ \cr
$K_{\Sigma\Sigma}^{(1)} = +0.86 fm$ & & & $K_{\Sigma\Sigma}^{(1)} = +0.27 fm$
\cr
$K_{\Sigma\Lambda}^{(1)} = +0.51 fm$ & & & $K_{\Sigma\Lambda}^{(1)} = +1.54
fm$ \cr
$K_{\Lambda\Lambda}^{(1)} = +0.04 fm$ & & & $K_{\Lambda\Lambda}^{(1)} = -1.02
fm$ \cr}}$$
\hrule
\vglue 1.3truecm

In the analysis of the low--energy data collected
in the past on these processes, one of the main difficulties
comes from the large spread in momentum of the
typical low--energy kaon beams, for $K^\pm$'s because of the degrading in a
``moderator'' of the higher--energy beams needed to transport the kaons
away from their production target, for $K_L$'s because of the large
apertures needed to
achieve satisfactory rates in the targets (typically bubble chambers): this
made unrealistic the proposals (advanced from the early seventies) of better
determining the low--energy
K--matrices by studying the behaviour of the cross sections for
$K^-p$--initiated processes at the
$\bar K^0n$ charge--exchange threshold$^{23}$. The high momentum resolution
available at DA$\Phi$NE will instead make such a goal a realistically
achievable one.

In this case one can no longer assume
$SU(2)$ to be a good symmetry of the amplitudes: under the (reasonable)
assumption that the forces are still $SU(2)$--symmetric, one can however
still retain the previous K--matrix formalism, but one can no longer
decouple the different isospin eigenchannels$^{24}$.
Introducing the orthogonal matrix {\bf R}, which
transforms the six isospin eigenchannels for $\bar KN$ ($I=0, 1$),
$\pi\Lambda$ ($I=1$ only) and $\pi\Sigma$ ($I=0, 1, 2$) into the
six physical charge channels $K^-p$, $\bar K^0n$, $\pi^0\Lambda$,
$\pi^-\Sigma^+$, $\pi^0\Sigma^0$ and $\pi^+\Sigma^-$, and the diagonal matrix
{\bf Q}$_c$ of the c.m. momenta for these latter, one can rewrite
the T--matrix for the $S$--waves in the isospin--eigenchannel space as
$$ \hbox{\bf T}_I^{-1} = \hbox{\bf K}_I^{-1} - i \hbox{\bf R}^{-1} \hbox{\bf
Q}_c \hbox{\bf R} \ , \eqno(24)$$
where {\bf K}$_I$ is a box matrix with zero elements between channels of
different isospin, and {\bf R}$^{-1}$ {\bf Q}$_c$ {\bf R} is of course no
longer diagonal.

Apparently this involves one more parameter, since it
also contains the element $K_{\Sigma\Sigma}^{(2)}$: in practice, if one is
interested in the behaviour of the cross sections in the neighbourhood of
the $\bar KN$ charge--exchange threshold, one can take the
c.m. momenta in the three $\pi\Sigma$ channels as equal, so that the
$I=2$, $\pi\Sigma$ channel
decouples from the $I=0,\ 1$ ones, since the ``rotated'' matrix
{\bf R}$^{-1}${\bf Q}$_c${\bf R}
has now only two non--zero, off-diagonal elements, equal to
${1\over2} (q_0 - q_-)$ (where the subscripts refer to the kaon charges),
between the $I=0$ and $I=1$ $\bar KN$ channels, the diagonal ones being the
same as in the {\sl fully} $SU(2)$--symmetric case, if one substitutes for the
$\bar KN$ channel momentum $q$ the average over the two charge states
${1\over2} (q_0 + q_-)$. $K^{(2)}_{\Sigma\Sigma}$ would however be important
for describing accurate experiments on $\pi\Sigma$ and $\pi\Lambda$ mass
spectra in the unphysical region below the $\bar KN$ threshold without
recourse to the $SU(2)$--symmetry limit: but the state--of--the--art of our
understanding of wave--functions even for the lightest nuclei is not
such as to make these isotopic--symmetry--breaking corrections relevant.

\vglue 0.6truecm
\leftline{\bf 5. Low--energy $K^+$ scattering is important, too. }
\vglue 0.3truecm

Better information on the $S=+1$ system is also essential in several cases. We
limit ourselves to mention only two of the problems coming to our mind.
Isospin symmetry, as can be seen from the previous section, is an
essential ingredient in the phenomenological analysis of the $KN$ system,
apart from obvious mass--difference effects, apparent only in the close
proximity of the thresholds, which one can describe by modifying the
K--matrix formalism as outlined above$^{24}$.

One way to check isospin symmetry is to relate the amplitudes derived from
{\sl charged} kaon scattering to the data from $K_S$
regeneration. Since isospin relates the scattering of charged kaons on protons
to the regeneration on neutrons (and {\sl vice versa}), the test is better
performed on an isoscalar nuclear target, such as deuterium or $^4$He. We
should have indeed, apart from kinematical corrections and CP--violation
effects,
$$ T(K_Lp \rightarrow K_Sp) = {1\over2} [T(K^0p \rightarrow K^0p) - T(\bar
K^0p \rightarrow \bar K^0p)] =$$
$$= {1\over2} [T(K^+n \rightarrow K^+n) - T(K^-n
\rightarrow K^-n)] \eqno(25)$$
and
$$ T(K_Ln \rightarrow K_Sn) = {1\over2} [T(K^0n \rightarrow K^0n) - T(\bar
K^0n \rightarrow \bar K^0n)] =$$
$$= {1\over2} [T(K^+p \rightarrow K^+p) - T(K^-p
\rightarrow K^-p)] \ ;\eqno(26)$$
when we introduce these equalities in a nuclear scattering calculation, as in
{\sl e.g.} a Glauber model, all {\sl elastic} multiple scattering effects
should apply equally to both the right-- and left--hand sides of the
equalities for an isoscalar nucleus, protecting the identity from a large
fraction of the ``nuclear'' effects$^{25}$.

Such tests could have been possible up to now only at higher momenta,
where the opening of inelastic channels in the $S=+1$ systems
complicates calculations further: a test performed in the elastic region
of this system should make things simpler and clearer.

The second problem, related in many theoretical analyses to observations from
inelastic electron and muon scattering on nuclei, namely to changes in the
electromagnetic properties and in the deep--inelastic structure functions of
nucleons bound in nuclei with respect to the free ones, is the
``antishadowing'' effect observed at momenta around $800\ MeV/c$ for $K^+$
scattering on nuclear targets$^{26}$.
Conventional Glauber--model calculations$^{27}$ led to
expect a ratio $(2\sigma_A)/(A\sigma_D)$ slightly less than unity and
decreasing with both the kaon momentum and the target mass number $A$,
while the measured
values were larger than unity and increasing with momentum.
This led to think,
as an explanation of this and of the aforementioned
electromagnetic phenomena, of
a ``swelling'' of the bound nucleons with respect to free ones, in line with
some of the explanations put forward for the ``nuclear'' EMC effect, though at
a much higher energy scale$^{28}$.

New data have recently confirmed this trend$^{29}$,
but only for momenta higher than approximately $600\ MeV/c$; a
possibility coming to mind is that the opening of inelastic channels, such as
$\pi KN$ (or more simply quasi--two--body ones as $KN^*$ and $K^*N$),
might necessitate the
introduction of inelastic intermediate states absent in a conventional
Glauber--model calculation, phenomenon analogous to the need to introduce
inelastic diffraction in the intermediate steps of a multiple--scattering
formalism to explain diffractive
processes on nuclei at much higher energies:
thus the data would be just showing the opening of the
threshold for such a phenomenon, particularly visible in the $K^+$--scattering
case because of the extremely long mean--free path of this hadron in nuclear
matter (about $7\ fm$).

Measurements of the $K^+$ cross sections on different nuclei in DA$\Phi$NE's
kinematical region, where $K^+N$ interactions are purely elastic, should help
close the issue when compared with accurate Glauber--model
calculations$^{27}$.

We would like to close reminding the reader that information on the $S=+1,\
I=0$ channel in this energy region is coming {\sl entirely} from
extrapolations from higher--momen-tum data, since $K^+$--scattering (and
regeneration) data on deuterium are {\sl not} available at momenta lower than
about $300\ MeV/c$: at present we have only a generic idea about the order of
magnitude of the {\sl absolute value} of the $KN$ $I=0$ scattering length,
expected to be of the order of some times $10^{-2}\ fm$ from forward
dispersion relations and the lowest--momentum regeneration data$^{18-20}$.
An accurate measurement of the cross sections for $K^+$ {\sl incoherent}
scattering on deuterium, possible at
DA$\Phi$NE over a wide angular range, would thus give us the first {\sl
direct} measurement of this quantity.

\vglue 2.0truecm

\centerline{\bf REFERENCES AND FOOTNOTES }

\vglue 0.3truecm

\item{1.} $K^\pm p$ data: W.E. Humphrey and R.R. Ross: {\sl Phys. Rev.}
{\bf 127} (1962) 1; G.S. Abrams and B. Sechi--Zorn: {\sl Phys. Rev.} {\bf 139}
(1965) B 454; M. Sakitt, {\sl et al.: Phys Rev} {\bf 139} (1965) B719; J.K.
Kim: Columbia Univ. report {\sl NEVIS--149} (1966), and {\sl Phys. Rev. Lett.}
{\bf 14} (1970) 615; W. Kittel, G Otter and I. Wa\v{c}ek: {\sl Phys. Lett}
{\bf 21} (1966) 349; D. Tovee, {\sl et al.: Nucl. Phys.} {\bf B 33} (1971)
493; T.S. Mast, {\sl et al.: Phys. Rev.} {\bf D 11} (1975) 3078, and {\bf D
14} (1976) 13; R.O. Bargenter, {\sl et al.: Phys. Rev.} {\bf D 23} (1981)
1484. $K^-d$ data: R. Armenteros, {\sl et al.: Nucl. Phys.} {\bf B 18} (1970)
425.

\item{2.} TST Collaboration: R.J. Novak, {\sl et al.: Nucl Phys.} {\bf B 139}
(1978) 61; N.H. Bedford, {\sl et al.: Nukleonika} {\bf 25} (1980) 509; M.
Goossens, G. Wilquet, J.L. Armstrong and J.H. Bartley: {\sl ``Low and
Intermediate Energy Kaon--Nucleon Physics''}, ed. by E. Ferrari and G. Violini
(D. Reidel, Dordrecht 1981), p. 131; J. Ciborowski, {\sl et al.: J. Phys.}
{\bf G 8} (1982) 13; D. Evans, {\sl et al.: J. Phys.} {\bf G 9} (1983) 885; J.
Conboy, {\sl et al.: J. Phys} {\bf G 12} (1986) 1143. A good description of
the experiment is in D.J. Miller, R.J. Novak and T. Tyminiecka: {\sl ``Low and
Intermediate Energy Kaon-Nucleon Physics''}, ed by E. Ferrari and G. Violini
(D. Reidel, Dordrecht 1981), p. 251.

\item{3.} $K_Lp$ data: J.A. Kadyk, {\sl et al.: Phys. Lett.} {\bf 17} (1966)
599, and report {\sl UCRL--18325} (1968); R.A. Donald, {\sl et al.: Phys.
Lett.} {\bf 22} (1966) 711; G.A. Sayer, {\sl et al.: Phys. Rev.} {\bf 169}
(1968) 1045.

\item{4.} R.H. Dalitz and A. Deloff: {\sl J. Phys.} {\bf G 17} (1991) 289. See
also ref. 10 for a wider bibliography on this subject.

\item{5.} For conventions and kinematical notations we have adopted the same
as: G. H\"ohler, F. Kaiser, R. Koch and E. Pietarinen: {\sl ``Handbook of
Pion--Nucleon Scattering''} (Fachinformationszentrum, Karlsruhe 1979), and
{\sl ``Landolt--B\"ornstein, New Series, Group I, Vol. 9b''}, ed. by H.
Schopper (Springer--Verlag, Berlin 1983), which have become a ``standard''
for describing $\pi N$ scattering.

\item{6.} B. Tromborg, S. Waldenstr\"om and I. \O verb\o : {\sl Ann. Phys.
(N.Y.)} {\bf 100} (1976) 1; {\sl Phys. Rev.} {\bf D 15} (1977) 725; {\sl Helv.
Phys. Acta} {\bf 51} (1978) 584.

\item{7.} J. Hamiltom, I. \O verb\o\ and B. Tromborg: {\sl Nucl. Phys.} {\bf B
60} (1973) 443; B. Tromborg and J. Hamilton: {\sl Nucl. Phys.} {\bf B 76}
(1974) 483; J. Hamilton: {\sl Fortschr. Phys.} {\bf 23} (1975) 211.

\item{8.} R.J. Hemingway: {\sl Nucl. Phys.} {\bf B 253} (1985) 742. Older data
are even poorer in statistics: see ref. 17 for a comparison. See also, for
formation on bound nucleons, B. Riley, I.T. Wang, J.G. Fetkovich and J.M.
McKenzie: {\sl Phys. Rev.} {\bf D 11} (1975) 3065.

\item{9.} G.C. Oades: {\sl Nuovo Cimento} {\bf 102 A} (1989) 237.

\item{10.} A.D. Martin, B.R. Martin and G.G. Ross: {\sl Phys. Lett.} {\bf B
35} (1971) 62; P.N. Dobson jr. and R. McElhaney: {\sl Phys. Rev.} {\bf D 6}
(1972) 3256; G.C. Oades and G. Rasche: {\sl Nuovo Cimento} {\bf 42 A} (1977)
462; R.H. Dalitz and J.G. McGinley: {\sl ``Low and Intermediate Energy
Kaon--Nucleon Physics''}, ed. by E. Ferrari and G. Violini (D. Reidel,
Dordrecht 1981), p. 381, and the Ph.D. thesis by McGinley (Oxford Univ. 1979);
G.C. Oades and G. Rasche: {\sl Phys. Scr.} {\bf 26} (1982) 15; J.P. Liu:
{\sl Z. Phys.} {\bf C 22} (1984) 171; B.K. Jennings: {\sl Phys. Lett.}
{\bf B 176} (1986) 229; P.B. Siegel and W. Weise: {\sl Phys. Rev.} {\bf C 38}
(1988) 2221.

\item{11.} B.L. Roberts: {\sl Nucl Phys} {\bf A 479} (1988) 75c; B.L. Roberts,
{\sl et al.: Nuovo Cimento} {\bf 102 A} (1989) 145; D.A. Whitehouse, {\sl et
al.: Phys. Rev. Lett.} {\bf 63} (1989) 1352.

\item{12.} See the review by J. Lowe: {\sl Nuovo Cimento} {\bf 102 A} (1989)
167.

\item{13.} J.W. Darewich, R. Koniuk and N. Isgur: {\sl Phys. Rev.} {\bf D 32}
(1985) 1765; H. Burkhardt, J. Lowe and A.S. Rosenthal: {\sl Nucl Phys.} {\bf A
440} (1985) 653; R.L. Workman and H.W. Fearing: {\sl Phys. Rev.} {\bf D 37}
(1988) 3117; R.A. Williams, C.R. Ji and S. Cotanch: {\sl Phys. Rev.} {\bf D
41} (1990) 1449; {\sl Phys. Rev.} {\bf C 43} (1991) 452; H. Burkhardt and J.
Lowe: {\sl Phys. Rev.} {\bf C 44} (1991) 607. For radiative capture on
deuterons (and other light nuclei), see: R.L. Workman and H.W. Fearing: {\sl
Phys. Rev.} {\bf C 41} (1990) 1688; C. Bennhold: {\sl Phys. Rev.} {\bf C 42}
(1990) 775.

\item{14.} Y.S. Zhong, A.W. Thomas, B.K. Jennings and R.C. Barrett: {\sl Phys.
Lett.} {\bf B 171} (1986) 471; {\sl Phys. Rev.} {\bf D 38} (1988) 837 (which
corrects a numerical error contained in the previous paper).

\item{15.} See the review presented by B.R. Martin at the 1972 Ba\v{s}ko Polje
International School, published in: {\sl ``Textbook on Elementary Particle
Physics. Vol. 5: Strong Interactions''}, ed by M. Nikoli\v{c} (Gordon and
Breach, Paris 1975).

\item{16.} R.H. Dalitz and S.F. Tuan: {\sl Ann. Phys. (N.Y.)} {\bf 10} (1960)
307.

\item{17.} Y.A. Chao, R. Kr\"amer, D.W. Thomas and B.R. Martin: {\sl Nucl.
Phys.} {\bf B 56} (1973) 46.

\item{18.} A.D. Martin: {\sl Phys. Lett.} {\bf B 65} (1976) 346.

\item{19.} A.D. Martin: {\sl ``Low and Intermediate Kaon--Nucleon Physics''},
ed. by E. Ferrari and G. Violini (D. Reidel, Dordrecht 1981), p. 97.

\item{20.} A.D. Martin: {\sl Nucl. Phys.} {\bf B 179} (1981) 33.

\item{21.} R.H. Dalitz, J. McGinley, C. Belyea and S. Anthony: {\sl
``Proceedings of the International Conference on Hypernuclear and Kaon
Physics''}, ed. by B. Povh (M.P.I., Heidelberg 1982), p. 201.

\item{22.} G.K. Atkin, B. Di Claudio, G. Violini and N.M. Queen: {\sl Phys.
Lett.} {\bf B 95} (1980) 447; {\sl ``Low and Intermediate Energy Kaon--Nucleon
Physics''}, ed. by E. Ferrari and G. Violini (D. Reidel, Dordrecht 1981), p.
131; J. Antol\'\i{n}: {\sl Phys. Rev.} {\bf D 43} (1991) 1532.

\item{23.} See the discussion on this point by D.J. Miller in: {\sl
``Proceedings of the International Conference on Hypernuclear and Kaon
Physics''}, ed. by B. Povh (M.P.I., Heidelberg 1982), p. 215.

\item{24.} G.C. Oades and G. Rasche: {\sl Phys. Rev.} {\bf D 4} (1971) 2153;
H. Zimmermann: {\sl Helv. Phys. Acta} {\bf 45} (1973) 1117; G. Rasche and W.S.
Woolcock: {\sl Fortschr. Phys.} {\bf 25} (1977) 501.

\item{25.} See the talk by V.L. Telegdi, in: {\sl ``High--Energy Physics and
Nuclear Structure''}, ed. by D.E. Nagle, {\sl et al.} (A.I.P., New York 1975),
p. 289.

\item{26.} E. Piasetzsky: {\sl Nuovo Cimento} {\bf 102 A} (1989) 281; Y.
Mardor, {\sl et al.: Phys. Rev. Lett.} {\bf 65} (1990) 2110. Older data at
higher momenta are to be found in: D.V. Bugg, {\sl et al.: Phys. Rev.} {\bf
168} (1968) 1466.

\item{27.} P.B. Siegel, W.B. Kaufmann and W.R. Gibbs: {\sl Phys. Rev.} {\bf C
30} (1984) 1256; Ya.A. Berdnikov, A.M. Makhov and V.I. Ostroumov: {\sl Sov. J.
Nucl. Phys.} {\bf 49} (1989) 618; Ya.A. Berdnikov and A.M. Makhov: {\sl Sov.
J. Nucl. Phys.} {\bf 51} (1990) 579.

\item{28.} P.B. Siegel, W.B. Kaufmann and W.R. Gibbs: {\sl Phys. Rev.} {\bf C
31} (1985) 2184; G.E. Brown, C.B. Dover, P.B. Siegel and W. Weise: {\sl Phys.
Rev. Lett.} {\bf 60} (1988) 2723; W.B. Kaufmann and W.R. Gibbs: {\sl Phys.
Rev.} {\bf C 40} (1989) 1729; W. Weise: {\sl Nuovo Cimento} {\bf 102 A} (1989)
265; J. Labarsouque: {\sl Nucl. Phys.} {\bf A 506} (1990) 539; M. Mizoguchi
and H. Toki: {\sl Nucl. Phys.} {\bf A 513} (1990) 685.

\item{29.} J. Alster, {\sl et al.}: report {\sl TRI--PP--92--5} (Vancouver
1982), presented at the International Symposium on Hypernuclear and Strange
Particle Physics, Shimoda (Japan), 9 -- 12 December 1991.

\bye